\documentclass[print]{mn2e}
\usepackage{epsfig}
\usepackage{color}

\begin{document}
\title[A possible relation between flare activity in SLSNe and GRBs]{A possible relation between flare activity in super-luminous supernovae and gamma-ray bursts}

\author[Y.-W. Yu \& S.-Z. Li]{
Yun-Wei~Yu$^{1,2}\thanks{E-mail: yuyw@mail.ccnu.edu.cn}$ and
Shao-Ze~Li$^{1}$ \\
$^{1}$Institute of Astrophysics, Central China Normal University,
Wuhan 430079, China \\
$^{2}$Key Laboratory of Quark and Lepton Physics (Central China
Normal University), Ministry of Education, Wuhan 430079, China\\}

\date{Accepted XXX. Received YYY; in original form ZZZ}

\pubyear{2016}


\label{firstpage}
\pagerange{\pageref{firstpage}--\pageref{lastpage}}
\maketitle

\begin{abstract}
Significant undulations appear in the light curve of a recently
discovered super-luminous supernova (SLSN) SN 2015bn after the first
peak, while the underlying profile of the light curve can be well
explained by a continuous energy supply from a central engine,
possibly the spin-down of a millisecond magnetar. We propose that
these undulations are caused by an intermittent pulsed energy
supply, indicating an energetic flare activity of the central engine
of the SLSN. Many post-burst flares were discovered during X-ray
afterglow observations of Gamma-Ray Bursts (GRBs). We find that the
SLSN flares described here approximately obey the empirical
correlation between the luminosity and time scale of GRB flares,
extrapolated to the relevant longer time scales of SLSN flares. This
confirms the possible connection between these two different
phenomena as previously suggested.
\end{abstract}




\section{Introduction}

During the past ten years, a series of modern supernova surveys have
discovered an unusual type of supernovae with an absolute magnitude
at peak emission of $M_{\rm AB}<-21$, which are more luminous than
normal supernovae by a factor of $\sim10-100$ (Benetti et al. 2014;
Bersten et al. 2016; Chatzopoulos et al. 2011; Chen et al. 2016a;
Chomiuk et al. 2011; Chornock et al. 2013; Gal-Yam et al. 2009,
2012; Howell et al. 2013; Inserra et al. 2013, 2016; Kangas et al.
2016; Leloudas et al. 2012; Lunnan et al. 2013, 2016; McCrum et al.
2014, 2015; Nicholl et al. 2013, 2016; Ofek et al. 2007;
Papadopoulos et al. 2015; Quimby et al. 2007, 2011; Smith et al.
2007, 2016; Vreeswijk et al. 2014; Yan et al. 2016).

The total radiated energy of a typical superluminous supernova
(SLSN) is on the order of $\sim10^{51}$ erg. If this radiation is
mainly powered as usual by the radioactive chain $\rm
^{56}Ni\rightarrow^{56}Co\rightarrow^{56}Fe$, then an extremely
large amount (several to several tens of solar masses) of
radioactive $^{56}$Ni would be required. In principle, such a high
mass of $^{56}$Ni could be produced by core-collapse explosions of
very massive progenitors with a very large explosion energy (Umeda
\& Nomoto 2008; Moriya et al. 2010) or by disruption explosions of
very massive progenitors due to pair-production instability (Barkat
et al. 1967; Heger \& Woosley 2002; Gal-Yam et al. 2009). In both
cases, however, the corresponding high masses of supernova ejecta
would lead to a broad slowly-evolving supernova light curve, whereas
the observational light curves rise and often decline rapidly.
Moreover, for the pair-instability events, there is still
controversy whether they occur locally and if they are related to
some SLSNe or not (e.g. McCrum et al. 2014; Georgy et al. 2017).
Therefore, generally speaking, the radioactivity power scenario is
seriously challenged by the very high luminosity of SLSNe.

Alternatively, a powerful central engine is believed to play an
essential role in driving SLSN explosions and in powering their
emission, by an instantaneous and/or a long-lasting energy injection
into the explosion-ejected stellar envelope. To be specific, at the
initial time of some SLSN explosions, their central engines could
impulsively provide a great amount of energy to the supernova
ejecta, and lead the ejecta to have a very high initial velocity
corresponding to a kinetic energy on the order of $10^{52}$ erg. If
these explosions happen in dense, extended circum-stellar material
(CSM; e.g., stellar wind and some particular material clusters),
then the ejecta can be subsequently heated by the conversion of the
kinetic energy through shock interaction between the ejecta and
surrounding material (Smith \& McCray 2007; Chevalier \& Irwin 2011;
Moriya et al. 2011, 2013; Ginzburg \& Balberg 2012; Inserra et al.
2016). Such interaction-powered supernovae can usually be indicated
by narrow Balmer emission lines (Chatzopoulos et al. 2011). On the
contrary, for hydrogen-poor SLSNe, shock interaction could usually
be negligible, and the supernova emission is probably powered
directly by a long-lasting central engine. During the past few years, a remarkable number of light curves of
SLSNe, even including several hydrogen-rich ones without narrow
lines, have been successfully explained with a continuous energy
injection (Dessart et al. 2012; Inserra
et al. 2013, 2016; Nicholl et al. 2013, 2016; Howell et al. 2013;
McCrum et al. 2014; Wang et al. 2015; Dai et al. 2016; Lunnan et al.
2016; Bersten et al. 2016; Yu et al. 2017). This indicates that the
central engine could be the most viable and most common energy
source for most SLSNe, which makes these SLSNe very relevant to
another engine-driven phenomenon: gamma-ray bursts (GRBs). In principle, different energy
sources could coexist in some SLSNe (e.g. Wang et al. 2015a; Lunnan
et al. 2016).

In the
framework of the long-lasting energy injection model, it is
convenient to connect SLSN light curves with the temporal
behaviors of their central engines. Specifically, while a continuous
energy injection determines the basic profile of the SLSN light curves,
it can also be expected that some light curve undulations could be
caused by flare activity of the central engines. In this case, the
modeling of these light curve undulations could provide a special insight in
probing the nature of these SLSN engines and even their possible connection with GRB
engines.

\section{SN 2015bn}
Recently, Nicholl et al. (2016) presented the multi-wavelength
observational results of Type I SN 2015bn, where some significant
undulations appear in its light curve during the first $\sim 150$
days. The dataset they provided is extensive and detailed, which
enables us to constrain and even distinguish different SLSN models.
As investigated in Nicholl et al. (2016), the basic profile of the
bolometric light curve of SN 2015bn, excluding the undulation
components, could in principle be modeled by a scenario of that the
fast evolving peak emission is caused by ejecta-CSM interaction,
while the late-time slowly-decline emission is dominated by
$^{56}$Co decays. However, it could still be very difficult (if not
impossible) to simultaneously model the light curve undulations by
successive collisions of the supernova ejecta with some massive
shells that were expelled prior to the supernova explosion, where
the structure of progenitor and its mass-loss history must be
designed and tuned elaborately. Another possibility they discussed
is that this SLSN is powered as usual by a long-lasting central
engine, which is on the focus of this paper. In our opinion, while
the underlying smooth light curve is constructed by a continuous
energy injection, the light curve undulations of SN 2015bn are
probably associated with the late flare activity of its central
engine.

\section{Fittings and results}
\subsection{Basic equations}
Following Kasen \& Bildsten (2010), the evolution of the internal
energy $E_{\rm int}$ of a supernova ejecta can be determined by (Yu
et al. 2015)
\begin{eqnarray}
\frac{ d E_{\rm int}}{d t} =  L_{\rm in} - L_{\rm sn}- 4\pi R_{\rm
}^2 v_{\rm }p  ,\label{Eint}
\end{eqnarray}
where $t$ is the time, $L_{\rm in}$ is the energy injection rate,
$L_{\rm sn}$ is the supernova luminosity, $R_{\rm }$ and $v_{\rm
}={dR_{\rm }/ dt}$ are the radius and speed of supernova ejecta, and
$p$ is the pressure that can be related to the internal energy by
$p={1\over3}(E_{\rm int}/{4\over3}\pi R_{\rm }^3)$. The dynamical
evolution of the supernova ejecta is given by ${dv_{\rm }/ dt}={4\pi
R_{\rm }^2p/ M_{\rm ej}}$, where $M_{\rm ej}$ is the total mass of
ejecta.  Then, by considering of the heat diffusion in the supernova
ejecta (see Eq. 11 in Kasen \& Bildsten 2010), the bolometric
luminosity of the supernova can be roughly determined by the
following formula:
\begin{eqnarray}
L_{\rm sn}={cE_{\rm int}\over R_{\rm }\tau}\left(1-e^{-\tau}\right),
\end{eqnarray}
where $c$ is the speed of light, $\tau=3\kappa M_{\rm ej}/4\pi R^2$
is the optical depth, and $\kappa$ is the opacity. For $\tau\gg 1$
the above equation reads $L_{\rm sn}= cE_{\rm int}/( R\tau)$, while
$L_{\rm sn}= cE_{\rm int}/R$ for $\tau \ll1$ (e.g., see Kotera et
al. 2013).

\subsection{The underlying profile of light curve}
Continuous energy release
from a SLSN engine could be due to the spin-down of a millisecond
magnetar (Woosley et al. 2010; Kasen et al. 2010) or due to the
feedback of fallback accretion onto a magnetar or a black hole
(Dexter \& Kasen 2013). In literature, the former model was employed more widely than the latter.
The energy injection rate in the spinning-down magnetar model can be expressed by the spin-down
luminosity of the magnetar as
\begin{eqnarray}
L_{\rm in}(t)=L_{\rm in}(0)\left(1+t/t_{\rm sd}\right)^{-2},
\end{eqnarray}
where the initial luminosity $L_{\rm in}(0)$ and the spin-down
timescale $t_{\rm sd}$ are taken as free parameters when we
interpret the basic trends of the light curve. As shown in Figure 1
(see also Figure 19 in Nicholl et al. 2016), the light curve of SN
2015bn can well be profiled by the magnetar engine model. The
excess of the first data could be caused by the magnetar-driven
shock breakout emission (Kasen et al. 2016), while the drop of the
last three data is due to the leakage of high-energy photons after
the ejecta gradually becomes transparent (Wang et al. 2015). By
introducing the expressions of the initial spin-down luminosity
$L_{\rm in}(0)=10^{7}B_p^2P(0)^{-4}\rm erg~s^{-1}$ and the timescale
$t_{\rm sd}=2\times10^{39}B_p^{-2}P(0)^2$ s, we can derive the
dipolar magnetic field strength and the initial spin period of the
magnetar to be $B_{p}=6.4\times10^{13}$ G and $P(0)=2.3$ ms. The
parameter values obtained here are somewhat different from those
presented in Yu et al. (2017), because the detailed structure of the
light curve was not taken into account there.

For a comparison, we also consider that the continuous energy injection comes from an outflow
that is driven as the feedback (e.g. disk wind) of a fallback
accretion onto the central compact object (a magnetar or a black
hole). By assuming a direct proportion to the fallback accretion
rate, the outflow luminosity due to the accretion feedback can be
written as (Piro \& Ott 2011)
\begin{eqnarray}
L_{\rm in}(t)=L_{\rm in}(0)\left[\left(t/t_{\rm
accr}\right)^{-1/2}+\left(t/t_{\rm
accr}\right)^{5/3}\right]^{-1}\label{fallback}.
\end{eqnarray}
Here the accretion timescale $t_{\rm accr}$ can roughly be estimated
by a free-fall timescale of $\sim(G\rho_0)^{-1/2}$ (Dexter \& Kasen
2013), which gives a typical value of a few hundred to a thousand of
seconds for a typical stellar density of $\rho_0\sim300~\rm
g~cm^{-3}$. During such a short time, most of extractable energy of
the fallback material has been released and injected into the
supernova ejecta. This quickly injected energy would mostly be
converted into the kinetic energy of the ejecta, which leads to a
high velocity of about $3\times10^9\rm cm~s^{-1}$. This velocity is
much higher than the observed photosphere velocity of SN 2015bn of
$\sim9\times10^8\rm cm~s^{-1}$ (Nicholl et al. 2016), which somewhat
disfavors the accretion feedback model, although a precise
calculation of photosphere velocity is dependent on the specific
matter distributions in the ejecta. Meanwhile, the supernova light
curve is predicted to decrease too quickly to be consistent with the
data after the peak, if an ejecta mass of $16M_{\odot}$ (for
$\kappa=0.1\rm cm^2g^{-1}$) is taken properly to fit the peak time.
As an attempt to save this model, we tentatively introduce an
artificial mass of $8M_{\odot}$ of radioactive $^{56}$Ni to improve
the modeling of the late emission, as shown in Figure 2. In
principle, this putative nickel mass could be produced by
core-collapse explosion\footnote{The pair-instability supernova
model can not be employed here because of the pre-assumed presence
of the central compact object. In any case, this model would also
predict a too high ejecta mass.}  of a $\sim 100M_{\odot}$
progenitor (Umeda \& Nomoto 2008), if the explosion energy can reach
$>5\times10^{52}$ erg, which could be provided by the accretion
feedback. However, the extremely high mass of the progenitor would
still lead to a too heavy supernova ejecta, the mass of which is at
least higher than the CO core of the star of $\sim 40M_{\odot}$ even
if the hydrogen and helium envelopes have been lost completely. This
mass is too high to be consistent with the ejecta mass of
$16M_{\odot}$ inferred from the fitting.

Therefore, in comparison, the spinning-down magnetar model can
explain the underlying profile of the light curve of SN 2015bn more
naturally and more self-consistently than the accretion feedback
model. In the following calculations, we will only take the magnetar
model into account.

\begin{figure}
\centering\resizebox{\hsize}{!}{\includegraphics{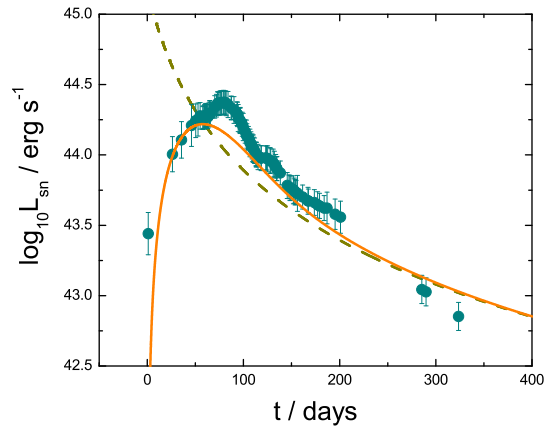}}
\caption{A fitting to the underlying profile of light curve of SN
2015b in the spinning-down magnetar model with parameters $L_{\rm
in}(0)=1.5\times10^{45}\rm erg~s^{-1}$ and $t_{\rm sd}=30 \rm day$
(solid line). The dashed line represents the energy injection rate.
The parameters for the supernova ejecta are adopted to
$\kappa=0.1\rm cm^2g^{-1}$, $M_{\rm ej}=10M_{\odot}$, and initial
velocity $v(0)=7.5\times10^8\rm cm~s^{-1}$. }
\end{figure}
\begin{figure}
\resizebox{\hsize}{!}{\includegraphics{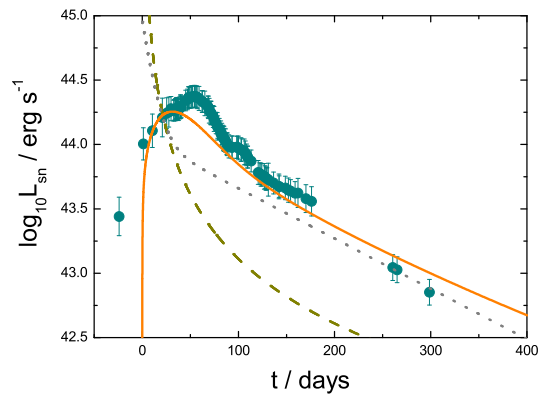}} \caption{A
tentative fitting to the underlying profile of light curve of SN
2015b by combining the heating effects due to accretion feedback and
radioactive decays of $^{56}$Ni (Solid line). The energy injection
rates due to accretion feedback and radioactivity are presented by
the dashed and dotted lines, respectively. The model parameters are
taken as $L_{\rm in}(0)=4.7\times10^{49}\rm erg~s^{-1}$, $t_{\rm
accr}=1000\rm s$, $\kappa=0.1\rm cm^2g^{-1}$, $M=16M_{\odot}$,
$v(0)=5\times10^8\rm cm~s^{-1}$ and $M_{\rm Ni}=8M_{\odot}$.}
\end{figure}

\subsection{Light curve undulations and flare activity}
A long-lasting central engine actually has been widely suggested to
account for many features of afterglow emission of GRBs. On one
hand, continuous energy release, usually from a spinning-down
millisecond magnetar, was always employed to explain the
shallow-decay or plateau afterglows of GRBs (Dai \& Lu 1998a,b;
Zhang \& Meszaros 2001; Yu et al. 2010; Rowlinson et al. 2013;
L{\"u} et al. 2015). On the other hand, more interestingly, a great
number of rapidly rising and declining X-ray flares have been widely
discovered in about one-third of Swift GRBs (Burrows et al. 2005;
Falcone et al. 2007; Chincarini et al. 2007; Wang \& Dai 2013; Yi et
al. 2016). This robustly indicates that many intermittent and
energetic activities have taken place on the GRB engines. By
considering of the high relevance between GRBs and SLSNe (Yu et al.
2017), it is natural to consider that the central engines of SLSNe
may also be able to make flare activity. As a result, some
undulations like those appearing in the light curve of SN 2015bn
could be caused by the delayed flaring energy release.

For an empirical fitting to the light curve of SN 2015bn, we
describe the energy release rate due to an engine flare by the
following formula:
\begin{equation}
L_{\rm flare}(t)=L_{\rm flare,p}\left[\left({t\over t_{\rm
flare,p}}\right)^{\alpha_1w}+\left({t\over t_{\rm
flare,p}}\right)^{\alpha_2w}\right]^{-1/w},\label{flares}
\end{equation}
which was usually adopted to fit the light curves of GRB X-ray
flares, where the structure parameters $\alpha_1$, $\alpha_2$, and
$w$ reflect the sharpness and smoothness of the flares. By assuming
a possible universal nature of flare activity, we take values of the
structure parameters refering to the fitting results of GRB flares
(Yi et al. 2016). Nevertheless, different from the GRB situation,
here we cannot directly confront Equation (\ref{flares}) with the
supernova light curve, because the flare energy can be completely
absorbed by the optical thick supernova ejecta. In other words,
engine flares can influence the supernova emission as pulsing energy
injections. Therefore, the consequent supernova light curves are
primarily related to the parameters $L_{\rm flare,p}$ and $t_{\rm
flare,p}$, but insensitive to the structure parameters. By
substituting expression (\ref{flares}) into Equation (\ref{Eint}),
we attempt to model the light curve undulations of SN 2015bn by
engine flare activity. Consequently, a perfect result is shown in
Figure 3, where three flares are invoked, which empirically
demonstrates the availability of the flare explanation of the light
curve undulations.

\begin{table}
\centering \caption{Flare Parameters}
\renewcommand{\arraystretch}{2.0}
\begin{tabular}{cccc}
 \hline
 \hline
 $L_{\rm flare,p}/10^{43}\rm {erg~s^{-1}}$&$t_{\rm flare,p}/\rm day$\\
\hline
 $32$ & $76$\\
  $5.5$ & $123$\\
   $1.5$ & $180$\\
\hline \hline
\end{tabular}
\end{table}

The parameters of the three SLSN flares are listed in Table 1. We
also plot these flares in the $L_{\rm flare,p}-t_{\rm flare,p}$
plane in Figure 4, where 200 GRB X-ray flares are presented for a
comparison. On one hand, for these GRB flares, an apparent
correlation of $L_{\rm flare,p}\propto t_{\rm flare,p}^{-1.27}$ was
found by Yi et al. (2016), although a large uncertainty of about two
orders of magnitude exists. On the other hand, as shown in Figure 4,
the SLSN flares have much longer timescales and lower luminosity
than the GRB flares.  In despite of these differences, we still find
that the three SLSN flares can basically fall into the 2$-\sigma$
uncertain region of the extension of the GRB flare correlation. Such
a consistency indicates that the central engines of SLSNe and GRBs
have a common behavior characteristics, which could be a natural
result of their same magnetar nature although the magnetic field
strengths of them are very different (Yu et al. 2017).

The statistics of GRB flares shows that the engines are usually more
active at earlier time. Then, one may query that why we can only
detect three very late flares at this first time and why we have
never yet detected any signature for more frequent earlier flares.
Our answer to these questions is,  even though the temporal
evolution of flares is very sharp, the undulations they can make in
supernova light curves are usually gentle, because of the trapping
of photons in the ejecta before the ejecta becomes completely
transparent. The earlier the time, the more serious the trapping
effect. As a result, any fluctuation in the energy release process
of an engine can in principle be smoothed in the consequent light
curve, in particular, before the photon diffusion timescale. In
other words, it is nearly impossible to detect any obvious signature
of early flares from a supernova light curve before its peak, no
matter how many flares have happened then. Nevertheless, in view of
the relatively lower magnetic fields of SLSN magnetars than GRB magnetars (Yu et al. 2017), it is still
reasonable to consider that the start time of the flare activity of SLSNe
is intrinsically later than that of GRBs.

\begin{figure}
\resizebox{\hsize}{!}{\includegraphics{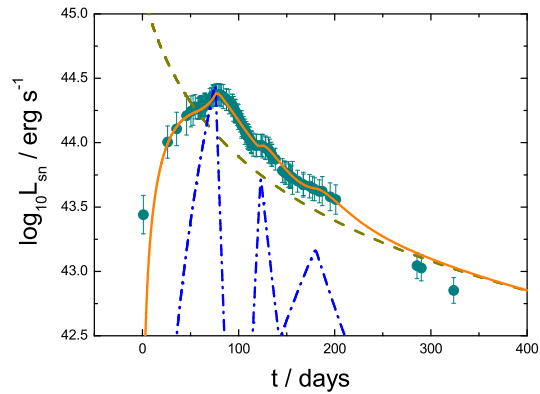}} \caption{Fittings
to the light curve of SN 2015bn (solid lines) in the spinning-down
magnetar model. For modeling the undulations, three flaring
intermittent energy injections are invoked (dash-dotted lines). The
parameters for the continuous energy injections are the same to
Figure 1.}
\end{figure}
\begin{figure}
\centering\resizebox{\hsize}{!}{\includegraphics{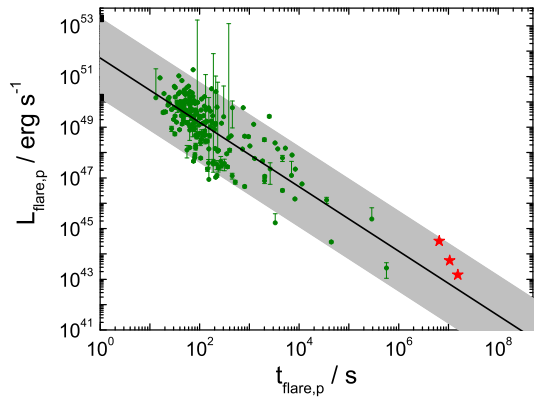}}\caption{A
comparison of the flares of SN 2015bn (stars) with 200 GRB flares
(solid circles) in the peak luminosity-peak time plane. The GRB data
is taken from Yi et al. (2016). The solid line presents the
correlation between the peak luminosities and peak times of these
GRB flares and the shaded region represents the 2$-\sigma$ uncertain
region of the relation.}\label{LC1}
\end{figure}

\section{Conclusion and discussions}
Recently, it has been suggested that Type I SLSNe and long GRBs
could be two branches from an united origin (Metzger et al. 2015; Yu
et al. 2017). To be specific, both are produced as outcomes of the
formation of a millisecond magnetar from core-collapse of a massive
rapidly rotating progenitor star. The primary difference between
them is the magnetic field strengths of their magnetars, which leads
to various observational differences between these two explosion
phenomena. The discovery of the light curve undulations of SN
2015bn, which is very fortunate, indicates that late but energetic
flare activity has taken place on its central engine. This strongly
reveals the similarity and connection between SLSNe and GRBs from a
new perspective, in view of the ubiquitous existence of GRB flares.
The possible universal $L_{\rm flare,p}-t_{\rm flare,p}$
relationship further indicates that the central engines of SLSNe and
GRBs could have a common nature, e.g., the magnetar nature as
supposed. This provides an independent and robust support to the
suggestion of Metzger et al. (2015) and Yu et al. (2017). The
different energy and time scales of the flares of SLSNe and GRBs
could arise from the different magnetic fields of their magnetar
engines.

In the framework of the magnetar engine model, we tend to believe
that the flare activity is associated with reconnections of
ultra-high multi-polar magnetic fields in the magnetar, by referring
to the magnetic reconnection model previously proposed by Dai et al.
(2006) for explaining GRB flares. Although the dipolar field derived
above for SN 2015bn is not so high, the discovery of some so-called
low-field magnetars in our galaxy (e.g. Rea et al. 2010) indicates
that the internal multi-polar fields of a magnetar could be much
higher than its surface dipolar field. In Dai et al. (2006), a
strong internal toroidal field is considered to form due to
differential rotation of the magnetar and subsequently float to the
stellar surface to be reconnected. Qualitatively, this scenario
could have many similarities with the physical processes of solar
flares. This was statistically confirmed by Wang \& Dai (2013), who
discovered that the distributions of energies, durations, and
waiting times of both GRB flares and solar flares can all be
understood within the physical framework of
self-organized-criticality avalanche-like processes. Nevertheless,
in Dai et al. (2006), the magnetar is assumed to simply consist of a
clearly-separated solid crust and core, which is probably invalid
for an extremely hot magnetar. An actual evolution of magnetic
fields of a newly-born magnetar could be much more complicated than
that considered in Dai et al. (2006). For example, fluid
instabilities could be involved (e.g. Cheng \& Yu 2014). Therefore,
an elaborate model is demanded to quantitatively describe the
magnetic reconnections and to account for the properties of the
observed SLSN flares as well as the related GRB flares.

An elaborate model is also needed to consolidate
our flare explanation for the light curve undulations and
distinguishing it from other possible scenarios (e.g., the CSM
interaction model).  First of all, the dynamics of the supernova
ejecta and the radiation transfer in it need to be described by a
more detailed radiation hydrodynamic code such as  the public
SuperNova Explosion Code (SNEC; Morozova et al. 2015). Furthermore,
in contrast to the 1D case considered here, a 2D simulation of Chen
et al. (2016) showed that the interaction between a magnetar wind
and a supernova ejecta can in fact lead to many fluid instabilities,
which mix the ejecta material and fracture the ejecta into
filamentary structure. The consequent inhomogenity and anisotropy of
the system could substantially influence the early dynamics and
emission of the supernova. It can even be expected that the clumpy
structure of the ejecta could also cause some light curve
undulations, which is worth to be investigated in future
simulations. In any case, it will be crucial and helpful to collect
more observational data to exhibit the details of early light curves
(in particular during the increasing phase) and to implement some
synergic multi-wavelength observations to SLSNe including in the
high-energy bands.

\section*{Acknowledgements}

The Authors thank M. Nicholl and F. Y. Wang for sharing the data of
SN 2015bn and GRB flares, respectively. This work is supported by
the National Natural Science Foundation of China (grant No.
11473008) and the Program for New Century Excellent Talents in
University (grant No. NCET-13-0822).









\end{document}